# An Enhanced Machine Learning-based Biometric Authentication System Using RR-Interval Framed Electrocardiograms

Song-Kyoo (Amang) Kim, Chan Yeob Yeun and Paul D. Yoo

**ABSTRACT** This paper is targeted in the area of biometric data enabled security by using machine learning for the digital health. The traditional authentication systems are vulnerable to the risks of forgetfulness, loss, and theft. Biometric authentication is has been improved and become the part of daily life. The Electrocardiogram (ECG) based authentication method has been introduced as a biometric security system suitable to check the identification for entering a building and this research provides for studying ECG-based biometric authentication techniques to reshape input data by slicing based on the RR-interval. The *Overall Performance* (OP) as a newly proposed performance measure is the combined performance metric of multiple authentication measures in this study. The performance of the proposed system using a confusion matrix has been evaluated and it has achieved up to 95% accuracy by compact data analysis. The *Amang ECG* (*amgecg*) toolbox in MATLAB is applied to the mean square error (MSE) based upper-range control limit (UCL) which directly affects three authentication performance metrics: the number of accepted samples, the accuracy and the OP. Based on this approach, it is found that the OP could be maximized by applying a UCL of 0.0028, which indicates 61 accepted samples within 70 samples and ensures that the proposed authentication system achieves 95% accuracy.

## I. INTRODUCTION
Recent new technologies in the mixed area between big data and artificial intelligence are changing the way of healthcare provision which could give the huge impact on the industry. This paper is targeted in the area of biometric data enabled security system based on the machine learning for the digital health. Biometric authentication is becoming more popular and is replacing traditional access control systems [1, 2]. The electrocardiogram (ECG) is one of the most recent biometric characteristics to be utilized in this manner [3, 4] because electrical conduction through the heat provides ECG data that can be applied to recognize individuals [5]. The usage of the ECG utilization as a biometric trait has been firstly introduced by the US military in 1977 [6]. Although much progress has been achieved since then, the challenges including data acquisition, pre-processing for data enhancement, and the assignment of authentication categories are still remained. In addition, the application of deep learning (DL) and other machine learning (ML) classification approaches become the part of the challenges in ECG based security systems [7]. ML is a subset of artificial intelligence that allows computers to perform tasks without explicit instructions and its techniques could design a verification model for authentication based on live ECG data [8-11]. ML algorithms construct a mathematical model based on data training for prediction (regressions), classification and pattern recognition [12]. The various ML applications include the analysis not only of videos, images, and sounds [13], but also of ECG data [11, 14-15].

ECG data can be sliced in the time domain to provide a signature that is unique to each individual [16-19] and time slicing with R-peak anchoring is highly effective for ECG analysis [16]. In this approach, only the R-peaks are used without detecting other ECG waveforms [16]. Slicing the ECG data based on the interval between heartbeats (the RR-interval) instead of the slicing window time is the basis of our proposed RR-interval framing (RRIF) method, wherein each sliced data sample is used as an input parameter for ML training.

Our novel authentication system uses a decision tree (DT) based multi-variable regression model for breaking down the dataset into smaller subsets to predict target values [20-21]. The convoluted nature of popular deep-structured ML means that such models also lack transparency and interpretability. Representative RRIF ECG data are shown in Fig. 1. We have updated the extended functions and demonstrations using the Amang ECG (*amgecg*) toolbox [19] for RRIF to generate the input for the regression approach during the training and testing phases.

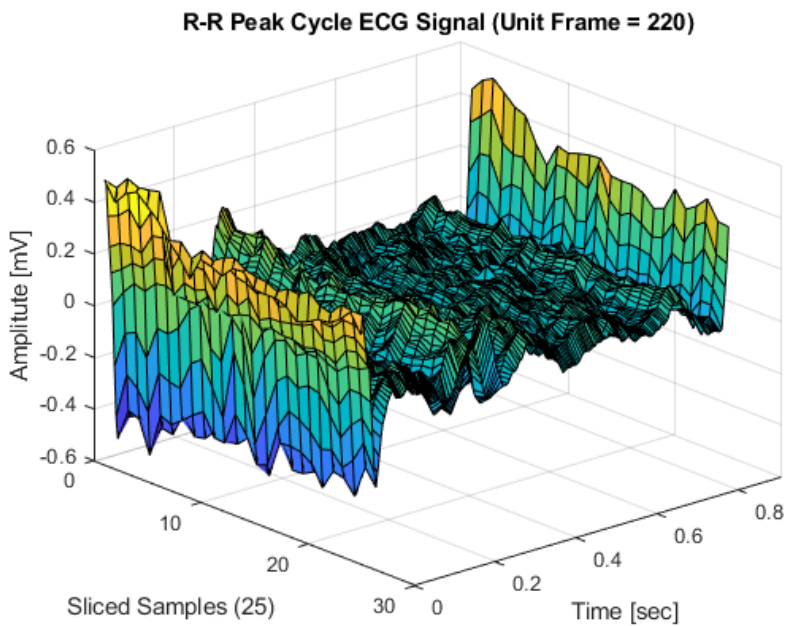

**FIGURE 1.** RR-Interval Framed ECG [22].

This rest of the article is divided into three sections. Section II describes the new ML-based authentication system based on the RRIF ECG input data, and evaluates its performance using a confusion matrix [23]. Section III suggests a new authentication performance measure which can reflect the quality of the ECG data or upper-range control limit (UCL), defined as the *Overall Performance* (OP). This is a single authentication performance measure that combines multiple conventional performance metrics. Section IV considers the three parameters that directly affect authentication performance and evaluates their relationship and impact. Finally, the new authentication system and its contributions are summarized in Section V.

## II. ELECTROCARDIOGRAM BASED AUTHENTICATION USING MACHINE LEARNING

This section shows the ECG-based authentication system using regression as a ML technique with RRIF ECG data that particularly complements the security check use case [19]. The sampling size of each RRIF is fixed at 220 by referencing from the standard ECG signal [24-26] and the system should thus identify the unknown entity.

### A. SEPUP OF EXPERIMENTS: SECURITY CHECK CASE

Several ML-based approaches have been tested to develop proper regression models, but our studies have indicated that the DT method provides the best performance with RRIF ECG data [16]. The performance has been evaluated between the DT (fine tree) and support vector machine (SVM) methods. The comparison results are shown in Table I and the DT-based regression method was selected for our new authentication system based on these results. The *Regression Leaner* function in MATLAB carried the comparison testing and the performance values in Table I are automatically created.

TABLE I

| | DT (Decision Trees) vs. SVM (Support Vector Machines) | |
|---|---|---|
| | DT (Fine Tree) | SVM (Fine Gaussian) |
| RMSE[1] | 0.10262 mV | 0.11995 mV |
| MAE[2] | 0.05634 mV | 0.06105 mV |
| Training Time | 563.06 s | 42573 s |

DT (preset=fine tree, min leaf size=4, surrogate decision splits=off) and SVM (present=fine Gaussian, kernel=Gaussian, kernel scale=.35, C=fitSVM, $\varepsilon$=fitSVM and standardization=true)

Therefore, the security check use case based on 60 ECG data samples is applied into this method. An ECG-based authentication system could not only identify employees but also exclude unknown individuals. It assumes that employees have registered their identities and the ECG data are reliable enough for both sessions of training and testing. For training, 60 samples were used to construct the dataset, 35 sourced from the PhysioBank database [22] and 25 from the Diabetes Complications Research Initiative [27]. The WFDB toolbox [28] is used for downloading and transforming the data from the PhysioBank [22] into MATLAB formats and the other data [27] were received already in MATLAB format. We randomly selected 10 additional samples (described as "unknown" data) from the same sources [22, 27] but the different sampling time periods were added during the testing phase [19].

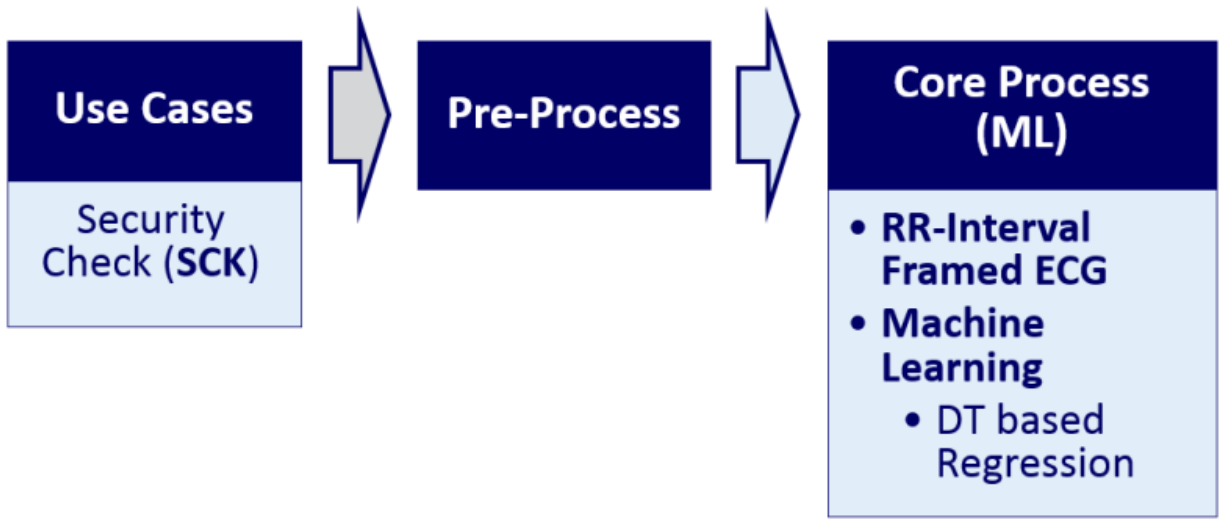

FIGURE 2. The process for the ML authentication system in the training phase [16, 19].

The authentication process initiated with use case categorization and the training dataset are pre-processed before starting the training phase. It is noted that the recommended pre-processing steps in our previous studies [16, 19] were fully applied because the data were sourced from medical equipment. The ECG data were collected at two different times, resulting in two datasets which were gathered from the hospitals. The ECG data were trained using the DT regression method (Fig. 2).

Although many different measures and techniques are cited in the biometric literature for feature selection filters [16], mutual information of which Shannon suggested [29] as the measure to score and rank the features were used in the DT model. Random variables which are represented as features and class labels have been applied for our mutual information theoretic feature selection [29]. A widely-used histogram-based approach to estimate the probability distribution has been applied [16]. This ignores the potential existence of more optimal subsets, comprising features that are not sequentially ranked in terms of their mutual information values. Hence, the entropy $H$ of a random variable $x$ with a probability mass function $p(x)$ could be defined as follows:

$$H(x) = -\oint_{\{x \in X\}} p(x) \cdot \log_2 p(x) dx \quad (1)$$

---

[1] RMSE: Root Mean Square Error
[2] MAE: Mean Absolute Error

where $X$ is a set of all possible outcomes of $x$. The concept of entropy gives rise to: (1) conditional entropy

$$\mathbb{E}[H(x|Y)] = -\oint_{\{x \in X, Y\}} p(x|y)p(y) \tag{2}$$

where the entropy $H$ of a probability function factor $x$ is conditional upon the knowledge of another random variable $Y$; and (2) mutual information,

$$I(x) = \oint_{\{x \in X, Y\}} p(x,y) \log_2 \frac{p(x,y)}{p(x)p(y)} d(x,y) \tag{3}$$
$$= H(x) - \mathbb{E}[H(x|Y)]$$

denoting the amount of information gained about $y$ as a result of knowing $x$. The decision function of the SVM, based on the sign of the distance from the separating hyperplane, is:

$$R_{emp}[f] = \frac{1}{l} \sum_{i=1}^{l} |f(x_i) - y_i| \tag{4}$$

The probabilistic bounds have the following form with probability at least $\eta$:

$$R[f] < R_{emp}[f] + \Phi\left(\sqrt{\frac{h}{l}}, \eta\right) \tag{5}$$

where $h$ is the capacity, $\Phi$ is an increasing function of $\frac{h}{l}$ and $\eta$, and $l$ is the number of examples.

The above equation indicates that the bounds involve a quantity measuring the complexity of the space, $l$ and $h$ of the function space. Therefore, the principle of empirical risk minimisation controls not only the empirical error but the model complexity [30]. The hyperplane decision function of the binary SVM with kernel method is:

$$f(x) = sgn\left(\sum_{i=1}^{l} y_i a_i \langle \Phi(x), \Phi(x_i) \rangle + b\right)$$
$$= sgn\left(\sum_{i=1}^{l} y_i a_i k(x, x_i) + b\right) \tag{6}$$

and the following quadratic program:

$$Max \quad W(a) = \sum_{i=1}^{l} a_i - \frac{1}{2} \sum_{i,j=1}^{l} a_i a_j y_i y_j k(x_i, x_j) \tag{7}$$

$$Subject\ To\ a_i \geq 0,\ i = 1, \ldots, l,\ and\ \sum_{i=1}^{l} a_i y_i = 0. \tag{8}$$

The number of unit frames between R peaks is 220 (the RR-interval) which is an average number in ECG grids [26]. However, the number of the frames in each RR-interval can be changed. The detailed process flow for the training and testing phases is summarized in Fig. 3, which has been adapted from our previous research [16].

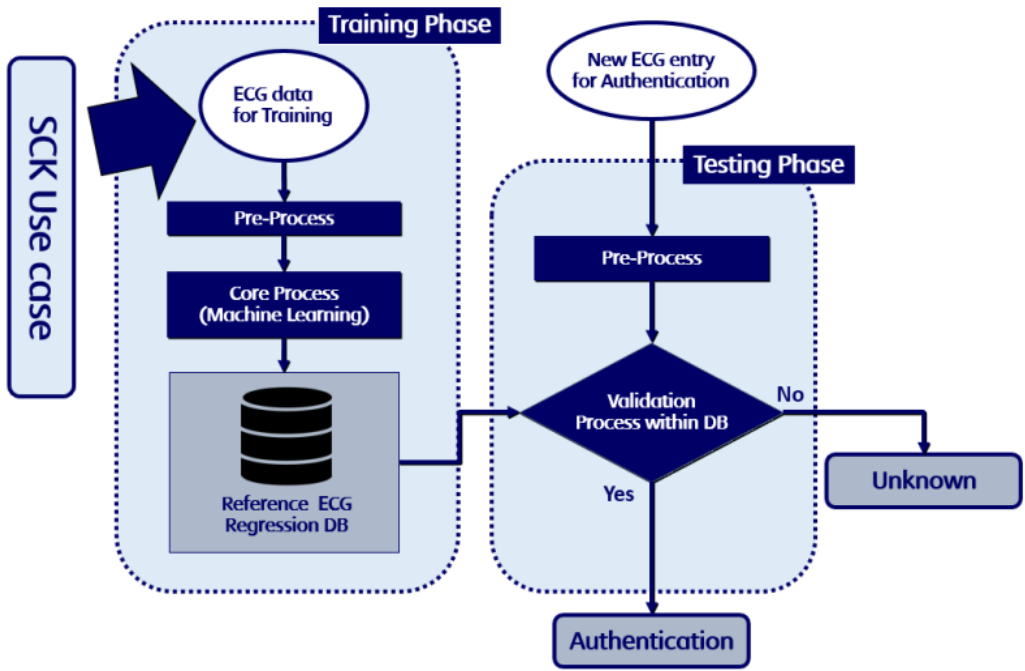

**FIGURE 3.** Process flow for ECG-based authentication in the SCK case [16].

The reference regression functions for each entity using the DT technique are generated and stored as a database in the training phase. During the testing phase, this database compares the ECG data when new ECG data are entered. The sampling time period for training the data was set to 50 s. The sampling time period in the testing phase shall be shorter because of the SCK (security check) properties in the use case category [16]. The core process builds reference regression functions for each set of trained ECG data and Fig. 4 shows some trained reference functions which make the ECG testing data be compared without modifying the sampling frequency.

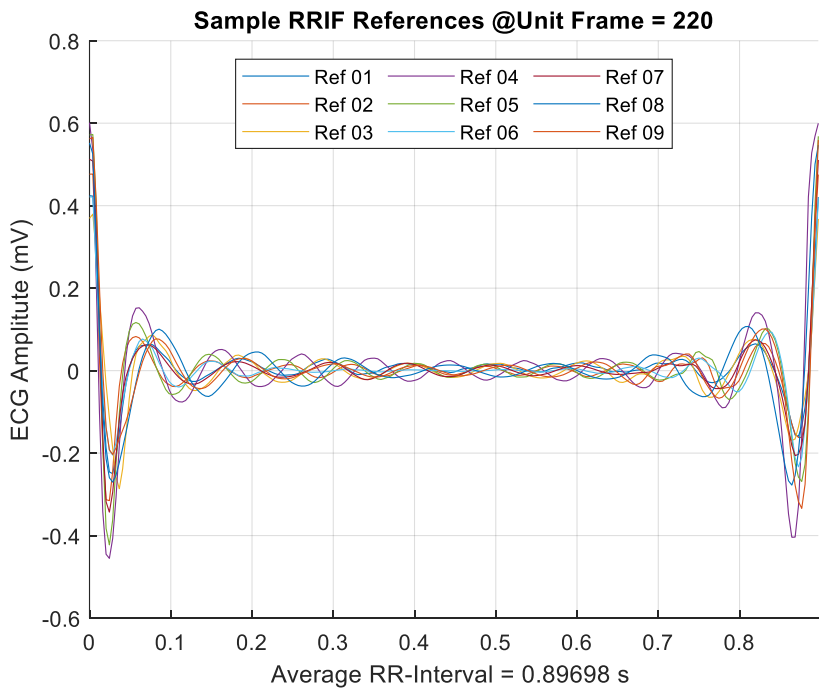

**FIGURE 4.** The reference ECG regression functions for the SCK case.

## B. EXPERIMENTAL RESULTS

Authentication performance was evaluated using a confusion matrix, i.e. a specific table layout that allows the performance of an algorithm to be visualized, typically a supervised learning based authentication [23]. Given the SCK use case, the ability to handle unknown entities is also required as part of the authentication process. Data quality measures based on the mean square error (MSE) are also used before detecting the testing ECG data. The first experiment uses 70 samples with a sampling time period of 15 s and it has been performed 100 times for the authentication testing. The confusion matrix is shown in Table II.

TABLE II
CONFUSION MATRIX FOR SECURITY CHECK CASE I

| 100 | | Actual ECG data | |
|---|---|---|---|
| | | Known | Unknown |
| Predicted ECG data | Known | 72 | 3 |
| | Unknown | 16 | 9 |

The authentication acceptance of the testing was achieved in 81 of the 100 experimental replicates (81%) and the unknown entity was identified nine times out of 12 (75%). According to the MSE based Upper Control Limit (UCL) [19], the accepted validation threshold could be increased based on the training data quality. When these higher data quality criteria were applied, we achieved the results shown in Table III. Several data quality measures could considered for the rejection criteria [19] and we selected the accurate percentage rate (APR) in our system. It is noted the values in a confusion matrix could vary even in the same number of accepted samples because the ECG testing data were randomly selected for each trial to make the security check (SCK) use case more realistic.

TABLE III
CONFUSION MATRIX FOR SECURITY CHECK CASE II

| 92 | | Actual ECG data | |
|---|---|---|---|
| | | Known | Unknown |
| Predicted ECG data | Known | 79 | 3 |
| | Unknown | 1 | 9 |

In this case, only 92 of 100 entities in the ECG testing data were accepted for validation using the reference regression function reflecting the data quality criteria. On the other hand, the accuracy of this biometric authentication system was 88 out of 92 experimental replicates (95.7%). Notably, 8% of ECG dataset (8/100) were excluded because they did not meet the criteria of the data quality [19].

## III. NEW PERFORMANCE MEASURE

Measuring the performance for ML algorithms include the evaluation of sample data quality and accuracy is essential part of any ML projects based on regression approaches [31]. The authentication system in this paper has two ML performance factors because the system adapts the data quality measures. When we compare the results of the two experimental setups (Tables II and III), it may not easy to tell which is better. Although the second approach is more accurate than the first (95.7% vs. 81%), it also has a higher sample rejection rate (eight samples were rejected before validating the data). To avoid confusion, we propose a new performance measure, the *Overall Performance* (OP), which is defined as Π in Equation (9):

$$\Pi = \left(\frac{\varphi}{N}\right) \cdot \chi, \qquad (9)$$

where $\varphi$ = the number of accepted data samples, $N$ = the total number of data samples and $\chi$ = the accuracy determined using the confusion matrix ($0 \leq \chi \leq 1$).

This new measure gives the overall performance of the authentication system based not only on accuracy but also on data quality. We calculated OP values for both of our experiments, resulting in OPs of 0.81 for the first experiment and 0.88 for the second. The authentication system with the data rejection criteria (second experiment) therefore achieves a better OP than the system without data rejection criteria (first experiment).

## IV. MSE BASED UCL DEPENDENCIES

Some ML performance measures depend on data quality criteria such as the UCL, which is based on the MSE and has a direct impact on the authentication performance. The UCL is a horizontal line on a control chart, which is typically placed at a distance of +3 standard deviations from the data mean [19].

We determined the relationship between the UCL and two performance measures: the accuracy of authentication and the number of accepted samples. The UCL-accuracy relationship is shown in Fig. 5, revealing that the best UCL range is 0.0032–0.0034 to achieve the highest accuracy (98.28%). The UCL range was determined by the training dataset, which was evaluated using the data quality functions in the *amgecg Toolbox* [19].

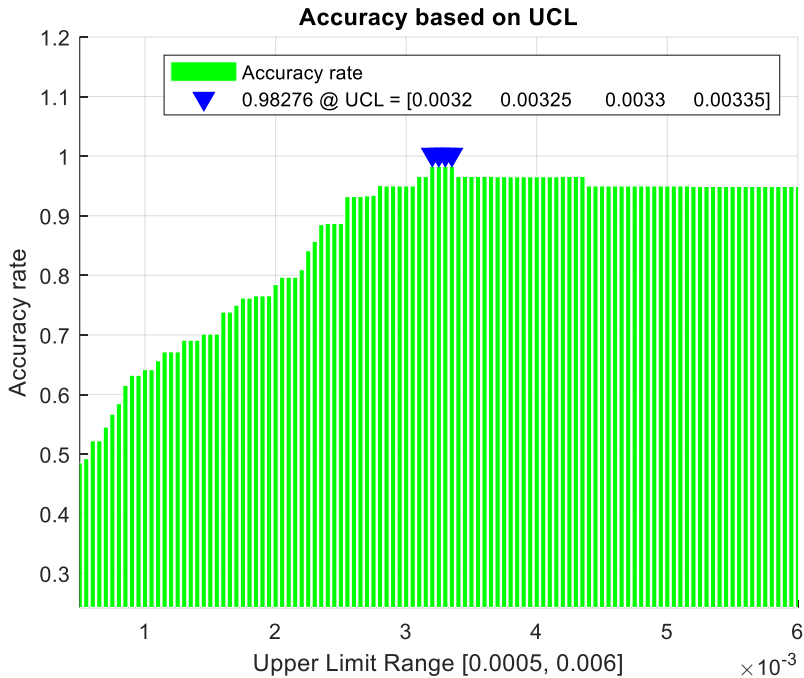

**FIGURE 5. Authentication accuracy based on the upper-range control limit.**

We also investigated the relationship between the UCL and the number of accepted samples (Fig. 6). Although the correlation is weaker than above, lower UCL values tend to be associated with larger numbers of accepted samples.

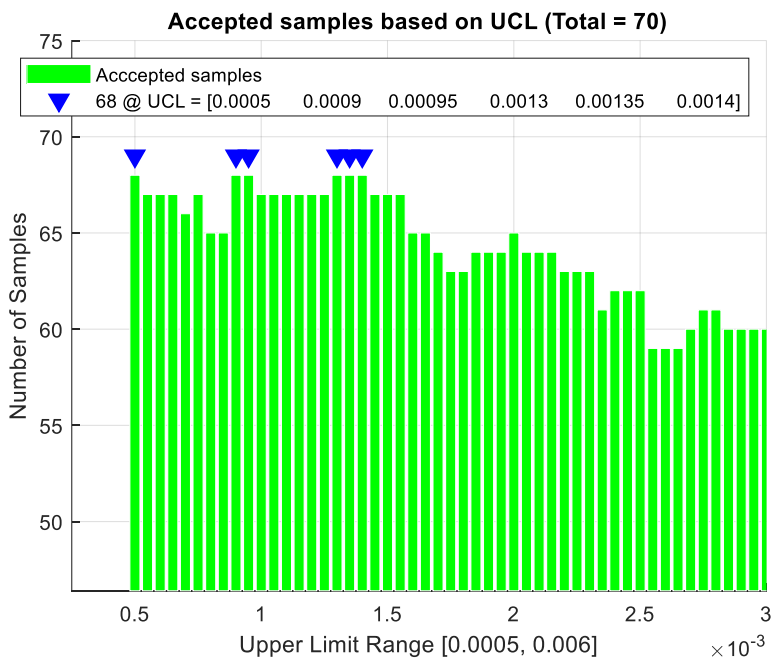

**FIGURE 6. Number of accepted samples based on the upper-range control limit.**

Alternatively, we can evaluate the relationship between the UCL and OP. We found that the optimal UCL based on accuracy was not the same as the optimal UCL based on the OP (Fig. 7). The optimal UCL associated with the best OP was 0.0028, which achieves 95% accuracy on the basis of 61 accepted samples from a total of 70.

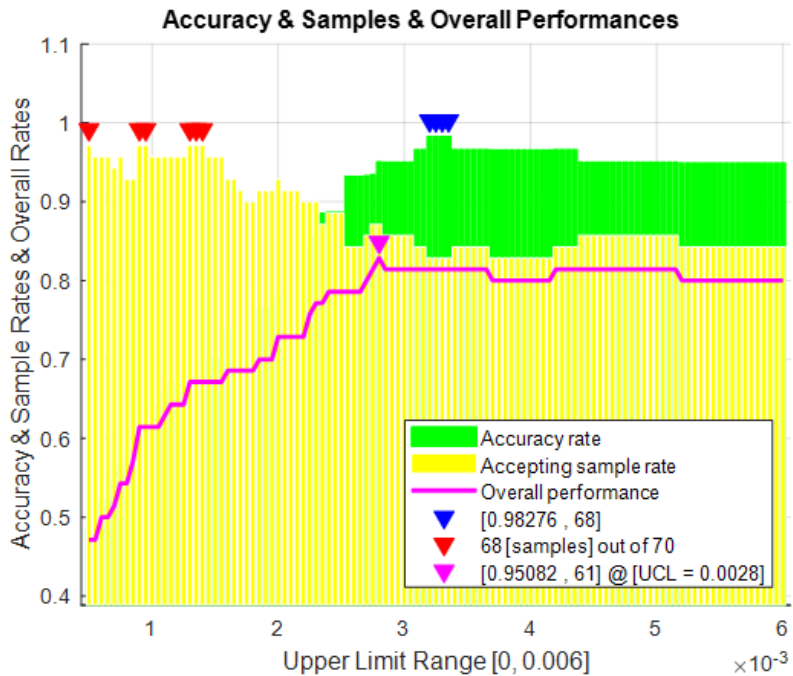

**FIGURE 7.** Authentication accuracy, the accepted sampling rate and OP based on UCL.

The optimal UCL may not be sustained in the same values if different datasets are adapted, but the experiments in this paper clearly demonstrated that optimal values for these parameters exist and could be adapted to improve the performance of authentication system.

## V. CONCLUSION

Biometric authentication systems have many advantages over traditional systems. However, the applications of ECG data depend on the particular use case and thus the nature of the authentication system could differ from other cases. A versatile RRIF biometric authentication system which uses a regression-based interpretable ML approach with the new OP measure based on the data quality (UCL) has been newly proposed for the security check use case. Total of 60 ECG data samples are trained to build the database for reference functions and each reference function for ECG data entities has been created by adapting a mutual information-based DT regression method. Additionally, the proposed security system performance was evaluated not only with a confusion matrix but also based on the overall performance measure. It is found that the OP could be maximized by using a UCL of 0.0028, which corresponds to 61 accepted samples out of 70 and 95% authentication accuracy.

## APPENDIX

This newly proposed biometric authentication system has been developed based on the *amgecg* toolbox [19] and the WFDB toolbox v0.10.0 [28]. The corresponding MATLAB codes are publicly available on GitHub[3] for users to do the demonstrations and users could fully understand the algorithms in this paper.

## ACKNOWLEDGMENT

Special thanks to the Guest Editor, Dr. Kim-Kwang Raymond Choo, who guides us to submit the proper topic of the journal. The authors acknowledge support from the Center for Cyber-Physical Systems, Khalifa University, under Grant Number 8474000137-RC1-C2PS-T3.

---

[3] https://github.com/amangkim/RRIFECGSec